\documentclass[onecolumn,amsmath,amssymb]{qm2008_abs}
\usepackage{graphicx}
\usepackage{dcolumn}
\usepackage{bm}

\newcommand{\beq}{\begin{equation}}
\newcommand{\eeq}{\end{equation}}
\newcommand{\bqa}{\begin{eqnarray}}
\newcommand{\eqa}{\end{eqnarray}}

\begin{document}

\title{{\Large Small Quarkonium states in an anisotropic QCD plasma }}
\bigskip
\bigskip
\author{\large Yun Guo}
\email{yun@fias.uni-frankfurt.de}
 \affiliation{Helmholtz Research School,
Johann Wolfgang Goethe Universit\"at,
Max-von-Laue-Str.\ 1, D-60438 Frankfurt am Main, Germany\\
Institute of Particle Physics, Huazhong Normal University, Wuhan
430079, China}
\bigskip
\bigskip

\begin{abstract}
\leftskip1.0cm
\rightskip1.0cm

We determine the hard-loop resummed propagator in an anisotropic
QCD plasma in general covariant gauges and define a potential
between heavy quarks from the Fourier transform of its static
limit. We find that the potential exhibits angular
dependence and that binding of very small quarkonium states is
stronger than in an isotropic plasma.
\end{abstract}

\maketitle

\section{Introduction}

Information on quarkonium spectral functions at high temperature
has started to emerge from lattice-QCD simulations; we refer to
ref.~\cite{Jakovac:2006sf} for recent work and for links to
earlier studies. This has motivated a number of attempts to
understand the lattice measurements within non-relativistic
potential models including finite temperature effects such as
screening~\cite{Mocsy:2005qw}. A detailed discussion of the
properties of the heavy-quark potential in the deconfined phase of
QCD is given in ref.~\cite{Mocsy:2007yj}, which also provides a
comprehensive list of earlier work.

In this paper, we consider the effects due to a
local anisotropy of the plasma in momentum space on the
heavy-quark potential. Such deviations from perfect isotropy are
expected for a real plasma created in high-energy heavy-ion
collisions, which undergoes expansion. We derive the HTL
propagator of an anisotropic plasma for general covariant gauges,
which allows us to define a non-relativistic potential via the
Fourier transform of the propagator in the static limit. We also
estimate the shift of the binding energy due to the anisotropy.

\section{Hard-Thermal-Loop self-energy and Propagator in an anisotropic plasma}

The retarded gauge-field self-energy in the hard-loop
approximation is given by~\cite{ThomaMrow}
\beq \Pi^{\mu \nu}(p)= g^2 \int \frac{d^3 {\bf k}}{(2\pi)^3} \,
v^{\mu} \frac{\partial f({\bf k})}{\partial k^\beta}
 \left( g^{\nu \beta} -
\frac{v^{\nu} p^\beta}{p\cdot v + i \epsilon}\right) \; .
\label{selfenergy1} \eeq
Here, $v^{\mu} \equiv (1,{\bf k}/|{\bf k}|)$ is a light-like
vector describing the propagation of a plasma particle in
space-time. The self-energy is symmetric,
$\Pi^{\mu\nu}(p)=\Pi^{\nu\mu}(p)$, and transverse,
$p_\mu\Pi^{\mu\nu}(p)=0$.

In a suitable tensor basis the components of $\Pi^{\mu\nu}$ can be
determined explicitly. For anisotropic systems there are more
independent projectors than for the standard equilibrium
case~\cite{Romatschke:2003ms}. To work in general covariant
gauges, we use a four-tensor basis developed in
ref.~\cite{dgs:plb} and the self-energy can now be written as $
\Pi^{\mu\nu}=\alpha A^{\mu\nu}+\beta B^{\mu\nu} + \gamma
C^{\mu\nu} + \delta D^{\mu\nu}$.

In order to determine the four structure functions explicitly we
need to specify the phase-space distribution function. We employ
the following {\em ansatz}:
\begin{equation}
f({\bf p}) = f_{\rm iso}\left(\sqrt{{\bf p}^2+\xi({\bf p}\cdot{\bf
n})^2} \right) ~,  \label{eq:f_aniso}
\end{equation}
where ${\bf n}$ is the direction of anisotropy in momentum space
and we choose ${\bf n}=(0,0,1)$ in this paper. The parameter $\xi$
is used to determine the degree of anisotropy. Thus, $f({\bf p})$
is obtained from an isotropic distribution $f_{\rm iso}(|\bf{p}|)$
by removing particles with a large momentum component along
$\bf{n}$.

Since the self-energy tensor is symmetric and transverse, not all
of its components are independent. We can therefore restrict our
considerations to the spatial part of $\Pi^{\mu\nu}$ and employ
the contractions in ref.~\cite{dgs:plb} to determine the four
structure functions. We do not list the rather cumbersome explicit
expressions for the four structure functions $\alpha$, $\beta$,
$\gamma$, and $\delta$ here since they have already been
determined in ref.~\cite{Romatschke:2003ms}.

The propagator
$i\Delta^{\mu\nu}_{ab}$ is diagonal in color and so color
indices will be suppressed.  In covariant gauge, its inverse is
given by
\begin{eqnarray}
&&\left(\Delta^{-1}\right)^{\mu \nu}(p,\xi)= -p^2 g^{\mu \nu}
+p^\mu p^\nu
-\Pi^{\mu\nu}(p,\xi)-\frac{1}{\lambda}p^\mu p^\nu \nonumber\\
&&\qquad\qquad\quad =(p^2-\alpha) A^{\mu\nu}+(\omega^2-\beta)
B^{\mu\nu} - \gamma C^{\mu\nu} - \delta
D^{\mu\nu}-\frac{1}{\lambda}p^\mu p^\nu
\end{eqnarray}
where $\omega\equiv p\cdot m$ and $m^{\nu}$ is the heat-bath
vector which equals to $(1,0,0,0)$ in the local rest frame.
$\lambda$ is the gauge parameter. Upon inversion, the propagator
is written as
\begin{equation}
\Delta^{\mu\nu} = \frac{1}{p^2-\alpha} \left[A^{\mu\nu} -
C^{\mu\nu}\right] +
\Delta_{G}\left[(p^2-\alpha-\gamma)\frac{\omega^4}{p^4}B^{\mu\nu}
+ (\omega^2-\beta)C^{\mu\nu} +
\delta\frac{\omega^2}{p^2}D^{\mu\nu}\right] -
\frac{\lambda}{p^4}p^\mu p^\nu ~,
\end{equation}
where
\begin{equation}
\Delta^{-1}_{G} = (p^2-\alpha-\gamma)(\omega^2-\beta) - \delta^2
\left[{\bf{p}}^2-(n\cdot p)^2\right]~,
\end{equation}
and the four vector $n^{\mu}=(0,{\bf n})$. For $\xi=0$, we recover
the isotropic propagator in covariant gauge.

\section{Heavy Quark Potential in an anisotropic plasma}

We determine the real part of the heavy-quark potential in the
nonrelativistic limit, at leading order, from the Fourier
transform of the static gluon propagator,
\begin{eqnarray}
V({\bf{r}},\xi) &=& -g^2 C_F\int \frac{d^3{\bf{p}}}{(2\pi)^3} \,
e^{i{\bf{p \cdot r}}}\Delta^{00}(\omega=0, \bf{p},\xi) \\
&=& -g^2 C_F\int \frac{d^3{\bf{p}}}{(2\pi)^3} \, e^{i{\bf{p \cdot
r}}} \frac{{\bf{p}}^2+m_\alpha^2+m_\gamma^2}
 {({\bf{p}}^2 + m_\alpha^2 +
     m_\gamma^2)({\bf{p}}^2+m_\beta^2)-m_\delta^4}~. \label{eq:FT_D00}
\end{eqnarray}
Here, $C_F$ is the color factor and the $\xi$-dependent masses $m_\alpha^2$, $
m_\beta^2$, $m_\gamma^2$ and $ m_\delta^2$ are given in
ref.~\cite{dgs:plb}.

\begin{figure}
\includegraphics[width=6.9cm]{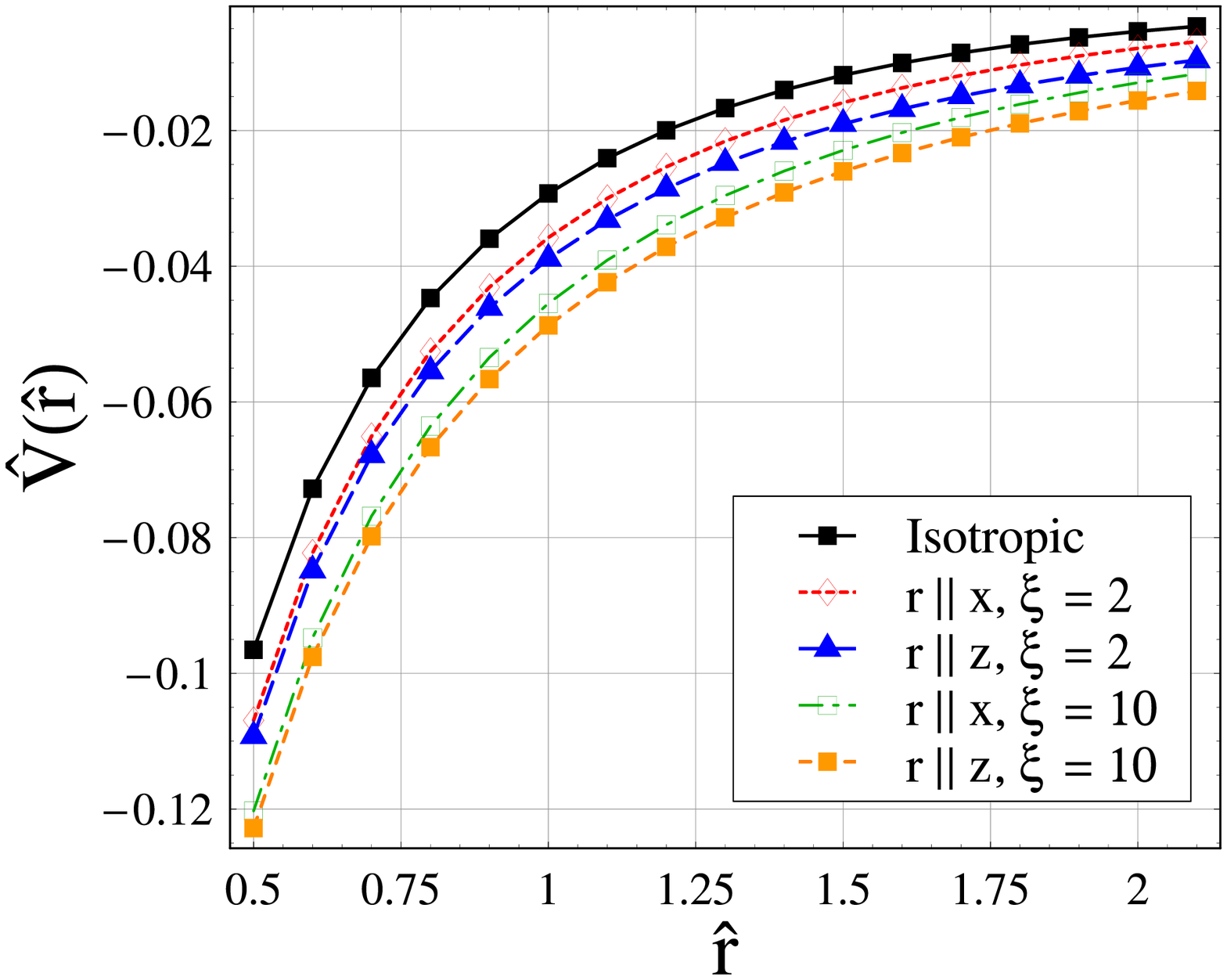}
\hspace{12mm}
\includegraphics[width=6.5cm]{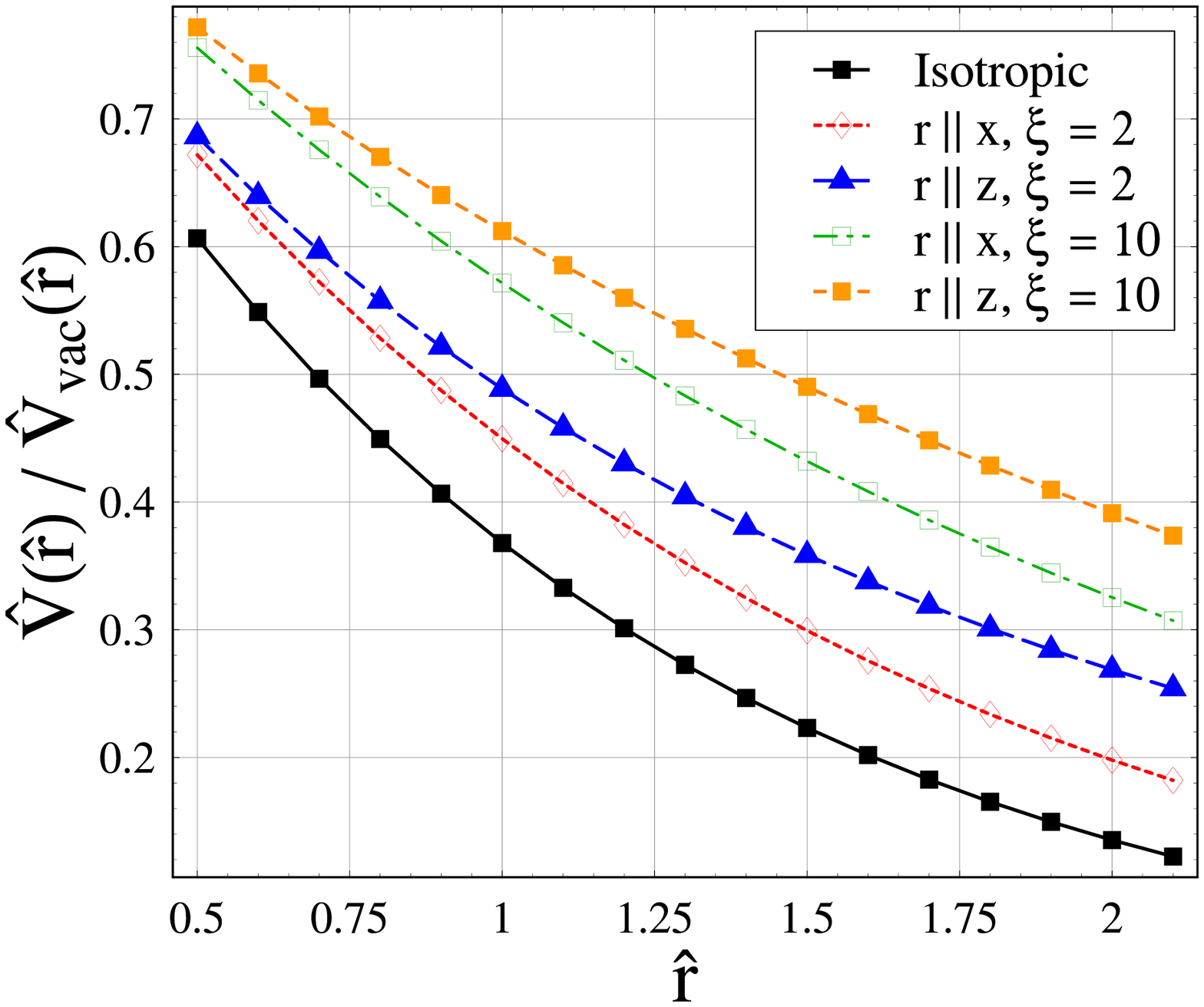}
\caption[a]{Heavy-quark potential at leading order as a function
of
  distance ($\hat{r}\equiv rm_D$) for ${\bf{r}}$ parallel to the
  direction ${\bf{n}}$ of anisotropy and ${\bf{r}}$ perpendicular to ${\bf{n}}$.
    \\ Left: the potential divided by the
  Debye mass and by the coupling, $\hat{V}\equiv V/(g^2 C_F m_D)$.
  Right: potential relative to that in vacuum.}
\label{fig:potential1}
\end{figure}

One may check some limiting cases~\cite{dgs:plb}. The isotropic Debye
potential is reproduced if $\xi=0$. On the other hand, when $r\to
0$ or $\xi\to \infty$, the potential then coincides with the
vacuum Coulomb potential. Generally, the integral
in~(\ref{eq:FT_D00}) has to be performed numerically. The poles of
the function are integrable. They are simple first-order poles
which can be evaluated using a principal part prescription. In
Fig.~\ref{fig:potential1} we show the potential in the region
$\hat{r}\equiv rm_D\sim1$ for various degrees of plasma
anisotropy, where $m_D$ is the Debye mass. One observes that in
general screening is reduced, i.e.\ that the potential at $\xi>0$
is deeper and closer to the vacuum potential than for an isotropic
medium. This is partly caused by the lower density of the
anisotropic plasma. However, the effect is not uniform in the
polar angle. The angular dependence disappears more rapidly at
small $\hat{r}$, while at large $\hat{r}$ there is stronger
binding for $\bf{r}$ parallel to the direction of anisotropy.
Overall, one may therefore expect that quarkonium states whose
wave-functions are sensitive to the regime $\hat{r}\sim1$ are
bound more strongly in an anisotropic medium.

\section{Binding Energy for small states}

Based on the above results, one can determine the binding energy of
bound quarkonium states. In particular, we will concentrate on the
shift of the binding energy due to the medium. An analytic estimate
can be obtained from non-relativistic perturbation theory (to first
order) for states with a Bohr radius times Debye mass small as
compared to the anisotropy parameter $\xi$.

For weak anisotropy, $\xi\ll1$, and distances
$\hat{r}\ll1$, we expand the potential as
\begin{equation}
V({\bf{r}})\approx V_{vac}(r)+\alpha m_D -
4\pi\alpha \xi m_D^2 \int \frac{d^3{\bf{p}}}{(2 \pi)^3} \,
e^{i{\bf p\cdot r}}\,\frac{\frac{2}{3}-({\bf p\cdot
n})^2/{\bf{p}}^2}{({\bf{p}}^2+m_D^2)^2}~,
\label{vapp}
\end{equation}
where $\alpha\equiv g^2 C_F/4\pi$.  If $m_D/\alpha m_Q\ll1$, where
$m_Q$ is the quark mass, the vacuum potential dominates and we can
use Coulomb wavefunctions to calculate the expectation value of
the perturbation given by the last two terms in the equation
above. Finally, the binding energy for such small-size states can
be expressed as
\begin{eqnarray}
E_{bin}\approx E_{vac}+\alpha m_D-\frac{\alpha \xi m_D}{6}~.
\label{binding}
\end{eqnarray}
Note that $E_{vac}<0$.
The relative shift of the binding energy due to the presence of the
(weakly anisotropic) medium is therefore given by
\begin{eqnarray}
\frac{\delta E}{E_{vac}}\approx \frac{4m_D}{\alpha m_Q} \, \left[
-1+\frac{\xi}{6} + \cdots \right]~
\quad\quad\quad,~~~({\rm for}~\frac{m_D}{\alpha m_Q} \ll \xi \ll 1).
\label{shift}
\end{eqnarray}
The first term corresponds to the reduced binding due to screening by
the medium while the second term is the correction due to the non-zero
anisotropy. The restriction to $m_D/\alpha m_Q\ll\xi$ is necessary
because terms of order ${\cal O}(m_D/\alpha m_Q)$ inside the brackets
have been neglected.

The above analysis applies also to excited states, provided that
their size (in units of the Debye length) is small as compared to
$\xi$. The shift of the binding energy for the $n^{th}$
state is
\begin{eqnarray}
\frac{\delta E}{E_{vac}}\approx
\frac{4 m_D} {\alpha m_Q}\, n \, \left[-1+\frac{\xi}{6}+\cdots\right]~.
\label{shift_n}
\end{eqnarray}
However, it should be clear that for realistic cases (i.e.,
temperatures, quark masses and coupling constant), our above
assumption that $m_D/\alpha m_Q\ll \xi$ is too extreme. For
intermediate-size states and general $\xi$, we must solve exactly the 3d
Schr\"odinger equation with the anisotropic potential. This work is in
progress.

\section{Discussion and Outlook}

We have determined the HTL gluon propagator in an anisotropic
(viscous) plasma in covariant gauge~\cite{dgs:plb}. Its Fourier
transform at vanishing frequency defines a non-relativistic potential
for static sources. We find that, generically, screening is weaker
than in isotropic media and so the potential is closer to that in
vacuum. Also, there is stronger binding of the quark pairs in the
anisotropic system. Our results are applicable when the momentum of
the exchanged gluon is on the order of the Debye mass $m_D$ or higher,
i.e.\ for distances on the order of $\lambda_D=1/m_D$ or less.

Following the discussion of the quark potential model in
ref.~\cite{Mocsy:2007yj}, at short distances, there is a string
contribution to the potential which is not calculable perturbatively.
However, at sufficiently high temperature, the perturbative contribution
dominates over the linear confining potential at the length scale
$\lambda_D$.  Roughly, this holds when $T$ is larger than $2T_c$.
In this case, our result is directly relevant for quarkonium
states with wavefunctions which are sensitive to the length scale
$\lambda_D$. Conversely, for those states whose length scale is
larger, one should sum the medium-dependent contributions
due to one-gluon exchange {\em and} due to the
string~\cite{Mocsy:2007yj}.

At very short distances, the contribution from the string can also be
neglected and one is dealing with perturbed Coulombic states.
However, for charmonium or bottomonium, the
string contribution is in fact important and it will be interesting to
determine their wave functions and binding energies
from a
potential which is a combination of our anisotropic potential and the
string contribution.

\section*{Acknowledgments}
The author gratefully acknowledges the collaboration with A.~Dumitru
and M.~Strickland and thanks the Helmholtz foundation, the Otto Stern
School at Frankfurt university and Quark-Matter 2008 for financial
support.


\vspace*{0.5cm}


\noindent
\begin{thebibliography}{50}
\medskip

\bibitem{Jakovac:2006sf}
A.~Jakovac, P.~Petreczky, K.~Petrov and A.~Velytsky,
Phys.\ Rev.\  D {\bf 75}, 014506 (2007) [arXiv:hep-lat/0611017];\\
G.~Aarts, C.~Allton, M.~B.~Oktay, M.~Peardon and J.~I.~Skullerud,
Phys.\ Rev.\  D {\bf 76}, 094513 (2007) [arXiv:0705.2198
[hep-lat]].

\bibitem{Mocsy:2005qw}
A.~Mocsy and P.~Petreczky,
Phys.\ Rev.\  D {\bf 73}, 074007 (2006) [arXiv:hep-ph/0512156].

\bibitem{Mocsy:2007yj}
A.~Mocsy and P.~Petreczky,
Phys.\ Rev.\ D {\bf 77}, 014501 (2008) [arXiv:0705.2559 [hep-ph]].

\bibitem{ThomaMrow}
S.~Mrowczynski and M.~H.~Thoma,
Phys.\ Rev.\  D {\bf 62}, 036011 (2000) [arXiv:hep-ph/0001164].

\bibitem{Romatschke:2003ms}
P.~Romatschke and M.~Strickland,
Phys.\ Rev.\ D {\bf 68}, 036004 (2003) [arXiv:hep-ph/0304092].

\bibitem{dgs:plb}
A.~Dumitru, Y.~Guo and M.~Strickland,
Phys.\ Lett.\ B {\bf 662}, 37 (2008) [arXiv:0711.4722 [hep-ph]].


\end{thebibliography}
\end{document}